\documentclass[aip,graphicx,amsfonts]{revtex4-1}

\usepackage{graphicx}
\usepackage{color}

\draft

\newcommand{\Eqref}[1]{Equation~(\ref{#1})}
\newcommand{\eqref}[1]{Eq.\,(\ref{#1})}

\newcommand{\eqrngref}[2]{Eqs.\,(\ref{#1}--\ref{#2})}
\newcommand{\figref}[1]{Fig.\,\ref{#1}}
\newcommand{\Figref}[1]{Figure \ref{#1}}
\newcommand{\tabref}[1]{Tab.\,\ref{#1}}
\newcommand{\Tabref}[1]{Table \ref{#1}}
\newcommand{\secref}[1]{\S\ref{#1}}
\newcommand{\Secref}[1]{Section \ref{#1}}
\newcommand{\appref}[1]{Appendix \ref{#1}}
\newcommand{\p}{\mathcal{P}}

\newcommand{\f}{\overline{\mathcal{F}}}
\newcommand{\n}{\mathcal{N}}

\newcommand{\IL}{\overline{I}}
\newcommand{\JR}{J_{\text{GS}}}
\newcommand{\JL}{J}
\newcommand{\fc}{\overline{f_c}}
\newcommand{\pc}{\overline{p_c}}

\newcommand{\B}{\overline{B}}
\newcommand{\A}{\overline{A}}

\newcommand{\vx}{\overline{x}}
\newcommand{\vsigma}{\overline{\sigma}}
\newcommand{\vm}{\overline{m}}
\newcommand{\vI}{\overline{A}}

\begin{document}

\title{A Unified Method for Inference of Tokamak Equilibria and Validation of Force-Balance Models Based on Bayesian Analysis}

\author{G.~T.~von~Nessi}
\email[]{greg.vonnessi@anu.edu.au}

\author{M.~J.~Hole}
\affiliation{
Research School of Physical Sciences and Engineering,
The Australian National University,
Canberra ACT 0200, Australia
}

\author{the~MAST~Team}
\affiliation{
EURATOM/CCFE Fusion Association, Culham Science Centre,
Abingdon, Oxon, OX14 3DB, UK
}

\begin{abstract}
A new method, based on Bayesian analysis, is presented which unifies the inference of plasma equilibria parameters in a Tokamak with the ability to quantify differences between inferred equilibria and Grad-Shafranov force-balance solutions. At the heart of this technique is the new concept of \emph{weak observation}, which allows multiple forward models to be associated with a single diagnostic observation. This new idea subsequently provides a means by which the the space of GS solutions can be efficiently characterised via a prior distribution. The posterior evidence (a normalisation constant of the inferred posterior distribution) is also inferred in the analysis and is used as a proxy for determining how relatively close inferred equilibria are to force-balance for different discharges/times. These points have been implemented in a code called BEAST (Bayesian Equilibrium Analysis and Simulation Tool), which uses a special implementation of Skilling's nested sampling algorithm [Skilling, Bayesian Analysis \textbf{1}(4), 833--859 (2006)] to perform sampling and evidence calculations on high-dimensional, non-Gaussian posteriors. Initial BEAST equilibrium inference results are presented for two high-performance MAST discharges.
\end{abstract}

{52.55.-s, 52.55.Fa}

\maketitle

\section{Introduction}
Reconstruction of the equilibrium magnetic field geometry is of central importance in both the control and analysis of tokamak plasmas. Indeed, precise, open-loop control and reliable inference from many disparate diagnostics systems will be an absolute necessity for scientists seeking to tune and control future tokamak experiments to sustain so-called burning plasmas; at the core of both these needs lies equilibrium reconstruction.

Many advances have been made in the area of equilibrium reconstruction since the 1970s, most of which have focused on making solving the Grad-Shafranov (GS) equation a fast computation.\cite{lackner1976,lao1985,wesson2011,hutchinson2005} While these efforts have resulted in equilibrium reconstruction becoming a cornerstone of routine shot-analysis on all of today's major tokamak experiments, current equilibrium reconstruction codes still must intrinsically assume the structure of the GS equation to make their inference. Thus, effects of flow, pressure anisotropy, and other points of kinetic physics are inescapably ignored in these reconstructions.

Modern tokamak equilibrium reconstruction works by finding the GS solution which best fits a set of given magnetic diagnostic data. This methodology has the advantage that such GS fits can be quickly computed but obviously removes the possibility of exploring how an inferred equilibrium may differ from a GS solution to better fit the actual data. Thus, an ideal compliment to these existing GS solvers is a tool, implemented as a piece of software, by which the strength of the a priori GS constraint could be adjusted to gain an intuition and quantification of how much an inferred structure (i.e. fit to diagnostics with minimal constraints) can deviate from a best-fit GS solution. Bayesian analysis is an ideal methodology upon which such a tool can be constructed, as it provides a readily accessible means to place non-intrinsic (i.e. adjustable) a priori constraints on inference parameters. The advantage of this Bayesian tool over present methods would be two-fold: it could quantify how that inferred structure differed from a GS solution; and it could produce expectations, uncertainties and higher-order statistical moments on inferred parameters in a mathematically rigorous way. This second point encompasses the tool's ability to replicate the functionality of modern equilibrium reconstruction codes.

The reason why such a code has not been previously developed is due to the fact that the associated inference problem would be non-linear in nature over a model (inference) parameter domain of high dimension. Such high-dimensional problems pose intrinsic computational difficulties and can take significant time and computational resources to perform (see Ch. 29--30 in MacKay\cite{mackay2003} for a nice discussion on these points). Indeed, traditional GS solvers only consider diagnostic fits that are GS solutions, which are subsequently characterised by a handful of parameters (usually about five in total) reflecting 1D parameterisations of the poloidal current and kinetic pressure profiles. Without an intrinsic GS constraint, this new code requires a collection of model parameters that characterise the 2D toroidal current profile, in addition to the 1D poloidal current and kinetic pressure profiles, which significantly increases the number of degrees of freedom (i.e. dimensionality of model parameters) in the inference.  This paper introduces a software tool based on Bayesian inference that uses new methods which help to overcome the computational obstacles associated with a high-dimensional, non-linear equilibrium reconstruction, while preserving the advantages associated with being able to control the strength of the GS a priori constraint. The result is a code which has a unique set of features, which wholly complement those of today's state-of-the-art GS solvers and is called the Bayesian Equilibrium Analysis and Simulation Tool or BEAST.

The paper is structured as follows: in \secref{sec::Overview} a general overview of Bayes' formula is given in the context of diagnostic data analysis. Next, \secref{sec::ForwardModels} gives an explanation of the plasma and diagnostic models used in BEAST. \Secref{sec::ObservationSplitting} details the concept of weak observations: an idea that enables multiple forward models to be associated with a single diagnostic observation. This is followed, in \secref{sec::Prior}, by a discussion on the design of the a priori constraint placed toroidal plasma current model. Computational obstacles and the general methodology used to overcome them are discussed in \secref{sec::ComputationIntro}, with relevant technical details and benchmarks presented in  \appref{sec::Computation}. Results are presented in \secref{sec::Results} for two high-performance MAST discharges. Finally, \secref{sec::Conclusions} contains concluding remarks and discusses future research directions.

\section{Overview of Bayesian Diagnostic Analysis}\label{sec::Overview}
Bayesian diagnostic inference is the mathematical foundation of BEAST. Here, we will only give a brief summary of how to apply Bayesian ideas to diagnostic data and encourage the interested reader to look elsewhere\cite{vonnessi2012,svensson2008,sivia2006,jaynes2003} for more details. In Bayesian diagnostic inference (assuming independent diagnostic observations) the goal is to statistically infer a vector of model parameters, denoted $\vm$, given a vector of diagnostic data and associated uncertainties, $\vx$ and $\vsigma$ respectively. A cornerstone of Bayesian inference is the idea that it is impossible to perform inference without making some background assumptions\cite{jaynes2003,mackay2003}, and these will be denoted as $\vI$. The specific nature of the assumptions used in this research will be made clear throughout the paper. Given this notation, Bayes' formula can be written as
\begin{equation}
\p(\vm|\vx,\vsigma,\vI)=\frac{\left(\prod_i\p(x_i|\vm,\sigma_i,\vI)\right)\p(\vm)}{\p(\vx,\vsigma,\vI)},\label{eq1}
\end{equation}
where the notation `$|$' is read as \emph{given} and the `$,$' read as \emph{and}; e.g. $\p(a,b|c,d,e)$ would read as \emph{the probability of $a$ and $b$ given $c$, $d$ and $e$}. In this formula, each factor has a common name in Bayesian theory: $\p(\vm|\vx,\vsigma,\vI)$ is called the \emph{posterior} and is the probability distribution of model parameters given a set of diagnostic observations and background assumptions; $\p(x_i|\vm,\sigma_i,\vI)$ is the \emph{likelihood} for a particular diagnostic observation and represents the probability that a given configuration of model parameters and diagnostic uncertainties generated the associated observation; $\p(\vm)$ is called the \emph{prior} and is a probability distribution which contains \emph{a priori} information about the model parameters themselves; and $\p(\vx,\vsigma,\vI)$ is the \emph{evidence}: a normalisation constant ensuring the left-hand side of Bayes' formula integrates to unity. Intuitively, one can thing of the posterior as representing an informed state of knowledge, when a prior understanding is updated with an observation, which is embodied in the likelihood. By using a posterior of on observation as a prior for another observation, an iterative process is created by which an initially given prior is updated with any number of observations. Finally, the evidence can loosely be thought of as the relative conviction one has about the inference of model parameters: a larger evidence corresponds to a better pairwise agreement between the prior and all likelihood distributions, given a fixed number of diagnostics and fixed uncertainties. This final point will be detailed further later in the section.

If considering only one observation corresponding to $x_i$ and $\sigma_i$, one can write Bayes' formula as
\begin{equation}
\p(\vm|x_i,\sigma_i)\propto\p(x_i|\vm,\sigma_i)\p(\vm).\label{eq2}
\end{equation}
Note that the $\vI$ corresponding to the background assumptions has been dropped to simplify the notation; this convention will be carried out through the remainder of the paper. As $\vx$ and $\vsigma$ are given and thus assumed to be constant, $\p(\vx,\vsigma)$ is also constant, justifying the proportionality in \eqref{eq2}. The \emph{forward model}, $\f(\vm)$, is implicitly contained within $\p(x_i|\vm,\sigma_i)$ and relates an arbitrary configuration of model parameters to a given set of observations. In particular, $\f(\vm)$ is a deterministic mapping from the vector space of model parameters to the vector space of associated diagnostic predictions. These predictions are meant to reflect the actual diagnostic reading if the physical system is in a state corresponding to the vector input of the forward model. Thus, providing the forward model represent the true physics of the diagnostic, values of $\f(\vm)$ will be the diagnostic measurement.

Likelihoods are taken to be mappings of the form
\begin{equation}
\p(x_i|\vm,\sigma_i)=\n(x_i-\mathcal{F}_i(\vm),\sigma_i^2),\label{eq2a}
\end{equation}
where $\n$ is represents a Gaussian distribution over pair-wise independent variables. The first argument of the Gaussian distribution represents the mean vector, and the second argument reflects the entries in a diagonal covariance matrix. This likelihood form is ubiquitous in diagnostic analysis, as diagnostic observations are often associated with Gaussian distributions whose standard deviation corresponds to the given error of the diagnostic. The structure of the likelihood is further justified when noting that Gaussian distributions serve to maximise the Shannon entropy, when only diagnostic observations and uncertainties are known.\cite{jaynes2003,sivia2006} When viewing the likelihood as a mapping from data-vectors to probability distributions over model parameters, \eqref{eq2a} indicates that all such output distributions are invariant, modulo translation determined by the given diagnostic observations. For such likelihoods, one may think of the evidence as a measure of the pairwise consistency between all observations and the prior knowledge. This can be understood by thinking of two arbitrary, freely-translating, Gaussian probability distributions with fixed variances multiplied together. The integral of the result will decrease as the expectation of these distributions move away from each other and is maximised when the expectations are identical; i.e. the overlap between the two distributions directly reflects the consistency between both associated observations. This statement is quantified by relative size of the evidence.

Using the particular likelihood in \eqref{eq2a}, one has the following expression for the posterior:
\begin{equation}
\p(\vm|\vx,\vsigma)\propto\left(\prod_i\n(x_i-\mathcal{F}_i(\vm),\sigma_i^2)\right)\p(\vm);\label{eq3}
\end{equation}
and thus, the posterior is a distribution over the space of model parameter configurations. \Eqref{eq3} is the general representation by which sampling statistics can be used to construct moments of the posterior distribution, e.g. expectation values and errors for model parameters. Finally, the evidence is calculated by integrating the right-hand side of \eqref{eq3} over the entire domain of model parameter configurations.

\section{Plasma and Diagnostic Forward Models}\label{sec::ForwardModels}
MAST is a well-diagnosed machine with over 100 equilibrium magnetic diagnostics and a Motional-Stark Effect (MSE) polarimetry system. This paper will focus on inferring the 2D toroidal current, along with the poloidal current and pressure profiles using equilibrium magnetics, MSE and Rogowski coil data measuring the total plasma current and toroidal field coil currents. 

\subsection{Toroidal Plasma Current Model}\label{sec::PlasmaModel}
\begin{figure}[h]
\includegraphics[width=.3\textwidth]{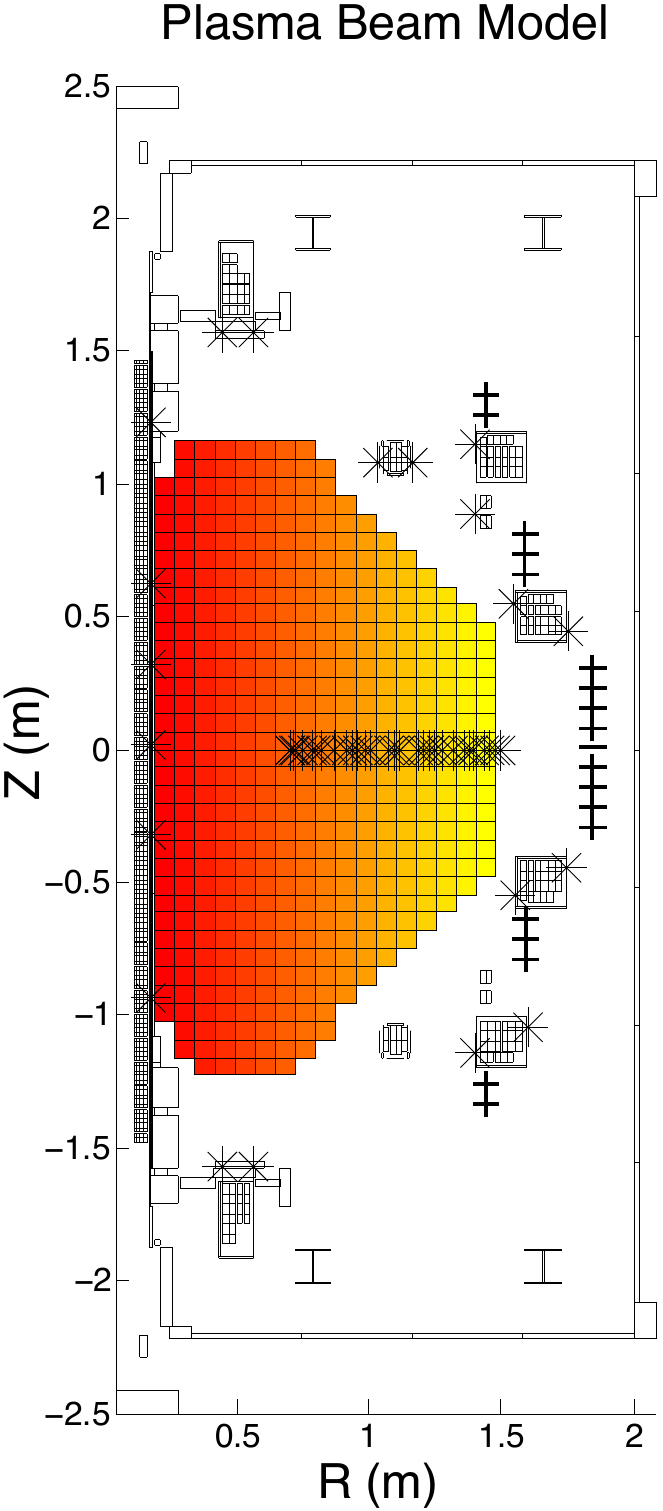}
\caption{\label{fig::PlasmaBeam}Colored rectangles indicate cross-sections of axisymmetric current elements designed to model the toroidal component of the plasma current. Conducting surface and poloidal field coil current beams are indicated by white rectangles having black outlines. Stars across the plasma midplane indicate MSE observation points, with stars outside the plasma volume corresponding to flux loop positions in the $(R, Z)$ plane. Finally, pickup coil positions and corresponding azimuthal orientations are indicated by thick black lines outside the plasma volume. One will note that there is a tightly packed configuration of vertically-oriented pickup coils in the central solenoid column and that outboard pickup coils are paired to form "crosses" with horizontal and vertical legs.}
\end{figure}

The toroidal plasma current is modelled by a collection of axisymmetric current beams of rectangular cross section, which fill out the plasma volume. This model was proposed in Svensson \& Werner\cite{svensson2008} and has been used successfully to infer the spatially-resolved toroidal plasma current in both the JET and MAST experiments \cite{svensson2008,hole2010,hole2011a,hole2011b,vonnessi2012}. \Figref{fig::PlasmaBeam} shows the position of plasma beams designed to model the toroidal plasma current. These beams have been selected such that their combined volume bounds the plasma volume for all the present MAST standard operational scenarios.\cite{vonnessi2012}

If one were to view the current through each beam in \figref{fig::PlasmaBeam} as an independent model parameter, the toroidal current model would correspond to 473 degrees of freedom. This number far exceeds the number of diagnostic observations available on any given MAST discharge, which is on the order of 150 including MSE measurements across the midplane. While this situation poses no fundamental problems in Bayesian inference, there is not enough diagnostic information to preclude unphysical ``screening'' solutions (i.e. where only the beams closest to the observation points are inferred with non-zero currents) from being favoured. One option available is to treat the plasma beams as nuisance parameters (model parameters which are integrated out in the final inference\cite{jaynes2003,sivia2006}) and extract only quantities derived off the toroidal currents which still maintain a physically meaningful inference. The other option is to utilise a physically informative prior (i.e. not a uniform distribution with bounds chosen so as to not preclude any physically realisable configuration for the associated parameter), and this will be discussed in \secref{sec::Prior}.

\subsection{Models for poloidal current and kinetic pressure}\label{sec::FAndPModels}
Using $R, Z, \phi$ to denote cylindrical coordinates with the $Z$ axis corresponding to the central axis of the tokamak, the poloidal flux can be related to the toroidal current by a standard application of the Biot-Savart law:
\begin{equation}
\psi(\IL; R, Z)=\frac{\mu_0}{2}\int_D\frac{\JL(R^\prime,Z^\prime)RR^\prime}{\sqrt{(Z-Z^\prime)^2+R^2+R^{\prime2}-2RR^\prime\cos\phi^\prime}}\,dR^\prime dZ^\prime d\phi^\prime,\label{eq::FAndP1}
\end{equation}
where $\IL$ denotes the collection of currents from the toroidal current model, $D$ the volume filled out by the plasma beams, and  $\JL(R,Z)$ being constructed as a 2D step function of the toroidal current density associated with the plasma model in \secref{sec::PlasmaModel}.\cite{vonnessi2012} Also, note that in \eqref{eq::FAndP1} the semicolon is preceded by model parameter inputs and followed by arguments which are given as fixed metadata in the inference.

Kinetic pressure, $p(\psi)$, and poloidal current, $f(\psi)$, are both taken to be polynomials of the poloidal flux, with $p(\psi)$ inferred with units of Pascals and $f(\psi)$ taken to represent a current, in Amperes, which is related to the toroidal magnetic field via 
\begin{equation}
f(\psi) = \frac{2\pi}{\mu_0}RB_\phi.\label{eq::FAndPModels1a}
\end{equation}
Using primes to denote derivatives, $p^\prime(\psi)$ and $f(\psi)$ are reconstructed from vectors of polynomial coefficients (denoted $\pc$ and $\fc$ for $p^\prime(\psi)$ and $f(\psi)$ respectively) according to
\begin{eqnarray}
p^\prime(\IL,\pc; R, Z) &=& p^\prime(\psi)\equiv p_{c,0}+p_{c,1}\psi+p_{c,2}\psi^2+p_{c,3}\psi^3;\label{eq::FAndPModels2}\\
f(\IL,\fc; R, Z) & = & f(\psi) \equiv f(\psi_\gamma) + f_{c,1}(\psi-\psi_\gamma)+f_{c,2}(\psi^2-\psi_\gamma^2)+f_{c,3}(\psi^3-\psi_\gamma^3),\label{eq::FAndPModels3}
\end{eqnarray}
where $\psi_\gamma$ is the poloidal flux at the last-closed flux surface and $f(\psi_\gamma)$ is the measured toroidal field coil current. The parameterisations in \eqrngref{eq::FAndPModels2}{eq::FAndPModels3} have been selected, as polynomial expansions of $f(\psi)$ and $p(\psi)$ have been validated using standard equilibrium reconstruction codes on a wide range of tokamak devices. The kinetic pressure can be recovered by integrating \eqref{eq::FAndPModels2} and noting that $p(\psi_\gamma)=0$. In these parameterisations, $\psi_\gamma$ is inferred as a nuisance parameter with an un-constraining uniform prior. Thus, the inference is on the polynomial coefficients depicted in \eqref{eq::FAndPModels2} and \eqref{eq::FAndPModels3}. The priors on $\pc$ and $\fc$ are uniform distributions whose bounds are chosen so as to not constrain the inference on MAST discharges.

\subsection{Diagnostic Forward Models}\label{sec::DiagnosticModels}
The poloidal flux, $\psi$, $f$ and $\B$ are related according to\cite{wesson2011,svensson2008,vonnessi2012}
\begin{eqnarray}
B_R(\IL; R, Z) & = & -\frac{1}{R}\frac{\partial\psi}{\partial Z},\label{eq::DiagnosticModels1}\\
B_Z(\IL; R, Z) & = & \frac{1}{R}\frac{\partial\psi}{\partial R},\label{eq::DiagnosticModels2}\\
B_\phi(\IL, \fc; R, Z) & = & \frac{\mu_0f}{2\pi R}.\label{eq::DiagnosticModels3}
\end{eqnarray}
The above and \eqref{eq::FAndP1} enable the relation between plasma beam currents and magnetic diagnostic predictions to be explicitly expressed and numerically implemented.\cite{svensson2008,vonnessi2012} Denoting the pickup coil forward and flux loop forward models as $\f_P$ and $\f_F$ respectively, one has the relations\begin{eqnarray}
\f_P(\IL;R,Z,\theta) &=& B_R \cos (\theta) + B_Z \sin(\theta), \label{eq::DiagnosticModels4} \\
\f_F(\IL;R,Z) &=& \psi,  \label{eq::DiagnosticModels5}
\end{eqnarray}
where $\theta$ is the angle between a pickup coil's normal and the mid-plane.\cite{svensson2008,vonnessi2012} Predictions for the MSE system can also be related to $\IL$ and $\fc$ via the forward model
\begin{eqnarray}
\f_M(\IL,\fc; R,Z,\A)=\frac{A_0B_Z+A_1B_R+A_2B_\phi}{A_3B_Z+A_4B_R+A_5B_\phi},\label{eq::DiagnosticModels6}
\end{eqnarray}
where $\A$ is a constant vector corresponding to the particular MSE viewing geometry.\cite{hutchinson2005,wesson2011,svensson2008,vonnessi2012} Finally, the measured total plasma current provides an additional observation corresponding directly to
\begin{eqnarray}
\f_{TP}(\IL) & = & \sum_iI_{L,i}.\label{eq::DiagnosticModels7}
\end{eqnarray}

\subsection{Force-Balance Model}\label{sec::ForceBalanceModel}
The GS equation is a manifestation of the force-balance relation requiring that kinetic (isotropic) pressure balance out the Lorentz force in axisymmetric magnetic confinement devices and can be written as\cite{grad2005,bellan2008,wesson2011}
\begin{eqnarray}
J_\phi(R,\psi) = 2\pi R\cdot p'(\psi)+\frac{\mu_0}{2\pi R}f(\psi)f^\prime(\psi).\label{eq::ForceBalanceModel1}
\end{eqnarray}
Instead of using $\JL(R,Z)$ from \eqref{eq::FAndP1} as the toroidal current density, one can use \eqref{eq::ForceBalanceModel1} and \eqrngref{eq::FAndP1}{eq::FAndPModels3}, to indirectly construct another 2D toroidal current density function, call it $\JR(R,Z)$, from the model parameters introduce in \secref{sec::PlasmaModel} and \secref{sec::FAndPModels}:
\begin{eqnarray}
\JR(\IL,\pc,\fc,\psi_\gamma; R, Z) = \left\{\begin{array}{rl} 2\pi R\cdot p^\prime(\psi) + \frac{\mu_0}{2\pi R}f(\psi)f^\prime(\psi),\quad\psi\ge\psi_\gamma\\0,\quad\psi<\psi_\gamma\end{array}\right.,\label{eq::ForceBalanceModel2}
\end{eqnarray}
where the conditional is meant to enforce no current flowing outside of the last-closed flux-surface. If \eqref{eq::ForceBalanceModel1} accurately represents the true force-balance physics of the plasma, then $\JL(R,Z)=\JR(R,Z)$ for all points within the plasma volume. Thus, any of the diagnostic forward models in \secref{sec::DiagnosticModels} should yield the same predictions regardless of whether they use $\JL(R,Z)$ directly or $\JR(R,Z)$ as their toroidal current density input, for their respective magnetic field calculations. This indicates that a second set of forward models can be constructed, in addition to those presented in \secref{sec::DiagnosticModels}, for which the force-balance condition in \eqref{eq::ForceBalanceModel1} is intrinsically held to be true. Heuristically, this new set of forward models is constructed via the following sequence of reductions:
\begin{enumerate}
\item $\IL \mapsto \psi$ (Amp\'ere's Law);
\item $\psi, \fc, \pc \mapsto \JR(R,Z)$ (Grad-Shafranov);
\item $\JR(R,Z), \fc \mapsto \overline{B}$ (Biot-Savart);
\item $\JR(R,Z), \overline{B} \mapsto $ magnetics, total plasma current and MSE predictions ($\f_P,\ \f_F,\ \f_M,\ \f_{TP}$).
\end{enumerate}
In both sets of models, $B_\phi$ is constructed directly from $\fc$ according to \eqref{eq::DiagnosticModels3}, and the total plasma current associated with $\JR(R,Z)$ is predicted by
\begin{eqnarray}
\f_{TP}(\JR(R,Z)) & := &\int_{\text{LCFS}}\JR(R,Z)\,dRdZ,
\end{eqnarray}
where LCFS is meant to indicate the cross-sectional area within the LCFS. In \secref{sec::ObservationSplitting}, the technique of observation splitting will be presented which enables both sets of forward models to be used simultaneously in the BEAST inferences.

At this point, one may suggest that a parameterisation of $\psi(R,Z)$ be inferred as the base set of model parameters, instead of using the plasma beam model in \secref{sec::PlasmaModel}. However, inferring $\psi(R,Z)$ directly would render the pickup coils and MSE observations essentially unconstraining (i.e. ultimately putting little information into the inference), as these forward models are based on derivatives of $\psi(R,Z)$ (c.f. \eqrngref{eq::DiagnosticModels1}{eq::DiagnosticModels2}). In particular, only small, local perturbations in $\psi(R,Z)$ would be needed to match any diagnostic observation associated with a forward model that has the local magnetic field as an input. Thus, the lack of magnetic diagnostic observations densely distributed across the entire plasma cross-section, makes it impossible to infer the poloidal flux directly with any acceptable accuracy or precision, without the use of physically-dominating a priori constraints (e.g. only considering the space of GS solutions). While diagnostic systems are being developed that can provide densely-packed magnetic field measurements within the plasma (e.g. the 2D MSE system proposed by J. Howard\cite{howard2008}), such diagnostics are still new and not widely deployed, rendering direct inference of $\psi(R,Z)$ a future research endeavour to be pursued when more of these systems come online.

\section{Weak Observations}\label{sec::ObservationSplitting}
In the previous section, two general sets of diagnostic forward models were presented which differ only in how they construct the toroidal current density: one does so directly from the plasma beam model of \secref{sec::PlasmaModel}; the other uses the plasma beam model to reconstruct the poloidal flux, which is subsequently fed, along with other model parameters, into the GS equation to produce a derived toroidal current density. Neither set of models sufficiently constrain the equilibrium inference to rule out unphysical solutions with a minimally informative prior. Given that both these models should be equally valid and non-conflicting (if one accepts \eqref{eq::ForceBalanceModel1} and associated assumptions), it is desirable to design an inference scheme which is able to utilise both these models as constraining entities before resorting to making more informative priors. One way to do this is to simply construct a new set of forward models from the previous two sets:
\begin{eqnarray}
\mathcal{F}^*_{i}(\vm):=(\tilde a_i\mathcal{F}_{i}(\vm)+\tilde b_i\mathcal{F}_{{\text{GS}},i}(\vm))\n(\mathcal{F}_{i}(\vm)-\mathcal{F}_{{\text{GS}},i}(\vm),\tilde\sigma_i^2),\label{eq::ObservationSplitting1}
\end{eqnarray}
where $\mathcal{F}(\vm)$ and $\mathcal{F}_{\text{GS}}(\vm)$ represent the forward models which use $\JL(R,Z)$ or $\JR(R,Z)$ as the toroidal current density, respectively; $\tilde a_i$, $\tilde b_i$ and $\tilde\sigma_i$ are arbitrary constants, with $\tilde a_i+\tilde b_i = 1$. \Eqref{eq::ObservationSplitting1} reflects the assumption that both sets of forward models should have equal predictions for a given set of model parameters that embody a force-balance configuration. By adjusting the relative $\tilde a_i$, $\tilde b_i$ constants, one can adjust the relative impact one model has over another. For example, if one wanted wanted to impose a weaker degree of certainty in the GS models, then they would decrease the values of $\tilde b_i$ relative to $\tilde a_i$. On the other-hand, the value of $\tilde\sigma_i$ reflects the degree of certainty in which both models are believe to yield the same predictions. 

This method of combining two disparate forward models into one and subsequently using them in the likelihood, we call making a \emph{weak observation}. Indeed, while this method enables one to use two forward models with one set of observations, the formulation in \eqref{eq::ObservationSplitting1} may create additional degeneracies in the inference. However, in the current situation, the additional constraint provided by the formulation in \eqref{eq::ObservationSplitting1} is consistently found to greatly help reduce the space of highly probably model configurations with minimal informative priors. 

Finally, this technique is quite general in that any number of forward models embodying theoretical constraints over the same set of forward models can be combined in a manner analogous to \eqref{eq::ObservationSplitting1}. In the case of multiple equalities, the Gaussian in \eqref{eq::ObservationSplitting1} is simply replaced by a product of Gaussians, with each reflecting a particular equality.

\subsection{Interpretation of Pressure and the Current Density Discrepancy}\label{sec::currentDiff}
As BEAST does not intrinsically impose a GS constraint (i.e. folding $\JL(R,Z)=\JR(R,Z)$ into the plasma model itself), one can interpret the difference between $\JL(R,Z)$ and $\JR(R,Z)$ as function which characterises a metric quantifying how far away the physical inference is from the space of GS solutions. On the other hand, $\JL(R,Z)=\JR(R,Z)$ not being intrinsically enforced undermines a strict physical interpretation of the pressure. Indeed, as pressure is not directly constrained by physical measurements, it's only physical context in the inference is via the GS relation; but the inferred equilibria itself need not necessarily be a GS solution.

The above dichotomy is the cornerstone behind the design of the BEAST code, which gives BEAST a complementary functionality relative to other GS solvers: it can quantify how a solution differs from a GS fit at the expense of relegating the pressure to a nuisance parameter. To this end, $\Delta I_i$ is defined as the current difference associated with each beam in the plasma current model:
\begin{equation}
\Delta I_i:=\int_{\Omega_i}|\JL(R,Z)-\JR(R,Z)|\,dRdZ,\label{eq::currentDiff1}
\end{equation}
where $\Omega_i$ is the cross sectional domain associated with the $i$th plasma beam current. Using the plasma beam configuration in \figref{fig::PlasmaBeam} and the expression in \eqref{eq::currentDiff1}, one can naturally define a 2D step function approximation of the differences in current densities as
\begin{equation}
\Delta J(R,Z):=|\JL(R,Z)-\JR(R,Z)|.\label{eq::currentDiff1}
\end{equation}
It is the inference of $\Delta J(R,Z)$ (i.e. the difference between inferred structure and the space of GS solutions), which differentiates BEAST from GS solvers.

Finally, It is clear that $\Delta J(R,Z)$ is tantamount to an additive correction to the GS equation itself; and thus, one could infer model parameters that parameterise such a correction within the overall inference. However, this would require that physical model parameters (i.e. the ones not associated with the correction itself) be intrinsically constrained to the space of GS solutions. Given the semi-linear structure of the GS equation, computation of any GS solution takes several iterative steps and would be required for every sample of or exploratory step taken in the posterior distribution, if a GS constraint was intrinsically enforced. As each inference will require up to hundreds of thousands of evaluations of the posterior, such a constraint would require a huge amount of computational time/resources. BEAST avoids these difficulties by constraining a set of forward models according to the GS relation and then inferring $\overline{\Delta I}$, as opposed to constraining the solutions space of model parameters directly and inferring an additive correction.

\section{Prior Selection for the Toroidal Plasma Current Model}\label{sec::Prior}
As indicated in \secref{sec::PlasmaModel}, minimally informative priors will not prevent unphysical ``screening'' solutions from being inferred. In the past, priors which enforce a degree of smoothening across the collection of plasma beams have been used in plasma current tomography to mitigate this effect.\cite{svensson2008,vonnessi2012} These priors (the conditional auto-regressive prior in Svensson and Werner\cite{svensson2008} and the Gaussian process prior in von Nessi, et. al.\cite{vonnessi2012}) were selected primarily because they both are fundamentally Gaussian distributions. These priors, combined with linear forward models in the likelihoods, ensured that the associated posterior was also a Gaussian distribution, which subsequently could be analysed by fast, analytic computational methods. Unfortunately, the non-linear nature of the force-balance forward models used in BEAST immediately render the associated equilibrium inference beyond the scope of analytic inversion techniques.

As there is no prior that will reduce BEAST inferences to analytic computations, prior selection is made on the second priority of preserving the interpretation of the posterior evidence and $\overline{\Delta I}$, defined in \eqref{eq::currentDiff1}. Both of these quantities will obviously be dependent on the prior, and this must always be taken into account. However, when comparing an inferred evidence between two different inferences, it is only a requirement that the priors in both inferences be the same, to draw a meaningful comparison. Thus, it is natural to focus on preserving the interpretation of $\overline{\Delta I}$ as a proxy for quantifying the discrepancy from a best-fit GS solution. Naively, one can design a prior which favours configurations where $\overline{\Delta I} = 0$:
\begin{eqnarray}
\p(\IL, \fc, \pc, \psi_\gamma) & = & \n(\overline{\Delta I},\sigma_{*}^2\overline{\overline{\mathbf{1}}}),\label{eq::Prior1}
 \end{eqnarray}
 where $\overline{\overline{\mathbf{1}}}$ is the identity matrix, with $\sigma_*^2$ a scalar constant to be set that reflects the strength of the prior. It is clear that the distribution in \eqref{eq::Prior1} does indeed form a prior, as it lacks any dependence on diagnostic data.
 
The prior in \eqref{eq::Prior1}, ensures that the inference is biased toward model configurations satisfying the GS relation embodied in \eqref{eq::ForceBalanceModel1} and subsequent forward models. This is exactly what is desired as we wish $\overline{\Delta I}$ to reflect the distance from the physical inference to the \emph{best-fit} GS solution. Moreover, the strength of this prior is readily adjusted by changing the value of $\sigma_*^2$ and is the adjustment that determines how strongly GS force-balance is enforced in the associated equilibrium inferences.

In practice, the prior in \eqref{eq::Prior1} is very effective at precluding non-physical screening solutions from being inferred, even with very high values of $\sigma_*^2$ (i.e. being less informative). Indeed, typical ranges for $\IL$ for the plasma beam configuration in \figref{fig::PlasmaBeam} range from 0--10kA in typical MAST discharge, and setting $\sigma_*$ as high as 5kA still gives inferences which are physical. 

\subsection{Extension of the Force-Balance Prior}\label{sec::Prior2}
While the prior in \eqref{eq::Prior1} is an effective means to set a preference for force-balance inferences, the variance associated with the distribution still has to be arbitrarily set. However, given that physical inferences are still to be had for a wide range of $\sigma_*^2$, it makes sense to extend to force-balance prior by making its variance a hyper-parameter: a model parameter to be inferred which parameterises the models/constraints used in the inference. Thus, \eqref{eq::Prior1} now becomes
\begin{eqnarray}
\p(\IL, \fc, \pc, \psi_\gamma,\sigma_*^2) & = & \n(\overline{\Delta I},\sigma_*^2\overline{\overline{\mathbf{1}}})\mathcal{U}_{[10^{-5},10]}(\sigma_*^2),\label{eq::Prior2}
 \end{eqnarray}
where $\mathcal{U}$ indicates a uniform distribution over the interval indicated by its subscript, for an associated model parameter reflected in its argument.

The advantage of using the prior in \eqref{eq::Prior2} is that the inferred value of $\sigma_*^2$ is now a scalar quantification of how far the inferred equilibrium is away from force-balance. Indeed, as smaller values of $\sigma_*^2$ serve to effectively reduce the degrees of freedom of our model parameters, the Bayesian inference will intrinsically favour smaller values of $\sigma_*^2$. Heuristically, this can be understood by noting that the Bayes' factor effectively penalises inferences which over-fit a given set of observations (see MacKay\cite{mackay2003} or Sivia \& Skilling\cite{sivia2006} for good discussions on this point).

\section{Computation}\label{sec::ComputationIntro}
In general, the posterior distribution for a MAST discharge will be a non-Gaussian distribution over approximately 500 model parameter dimensions. Integration of and sample generation for such high-dimensional posteriors historically have been computationally difficult and/or intractable problems.\cite{mackay2003,sivia2006} Moreover, traditional methods based on approximating the posterior using Gaussian distributions will generally give poor and/or unreliable results. Indeed, the only way such approximations can work is if all dimensional projections of the posterior are well-approximated by one or more Gaussians. While this may be reasonable to assume for a low dimensional problem, in many dimensions it will generally be the case that there will be at least some projections which are poorly approximated by normal distributions. To compound the problem, the complexity of the forward models often associated with so many model parameters makes deciphering which projections would be well-suited to such an approximation a difficult or impossible proposition. 

The other, common alternative to Gaussian approximation to analyse high-dimensional posterior distributions is to use algorithms based on the Markov Chain Monte-Carlo (MCMC) concept. There is a wealth of literature on the topic of MCMC algorithms that describe their associated issues in high-dimensional inference problems.\cite{mackay2003,sivia2006,jaynes2003} Putting these issues aside, MCMC methods can only yield samples of the posterior and are unable to provide an estimation of the posterior evidence $\p(\vx,\vsigma)$. Indeed, the issue of high-dimensional posterior integration is a topic of current research for which there are few viable algorithms.\cite{mackay2003,neal1998,neal1993}. 

Recently, Skilling\cite{skilling2006,sivia2006} developed the Nested Sampling (NS) algorithm to specifically calculate the evidence of Bayesian posterior distributions, which also provides a means by which posterior sampling can be simulated; this will be elaborated upon in what is to follow and in \appref{sec::Computation}. In general, NS works by transforming the multi-dimensional evidence integral to a one-dimensional integral that can be integrated via a statistical quadrature. This quadrature is constructed from samples taken from the \emph{prior} under a likelihood constraint. However, in addition to providing a means to calculating the evidence, this collection of samples can also be used as a compressed representation of the posterior itself. That is, a posterior probability may be canonically associated with each quadrature sample, so that probabilistic selection of elements from the entire set, based on this associated posterior probability, reflects direct sampling from the posterior. In summary, NS provides a means for both efficient evidence calculation and sample generation, with acceptable dimensional scaling; and thus, this algorithm was selected to form the backbone of BEAST.

Beyond these generalities, there are many subtleties and computational obstacles associated with the use of the NS algorithm. These points are discussed and addressed in \appref{sec::Computation}, along with some benchmarking results presented for different configuration of BEAST run parameters.

\section{Results}\label{sec::Results}
To demonstrate the capabilities of BEAST, two high-performance MAST discharges are analysed: \#22254 at 350ms and \#24600 at 265ms. Both of these discharges/times are strongly shaped double-null diverter (DnD) plasmas with 3.13MW and 3.35MW of NBI heating respectively. However, these discharges differ in that \#22254 at 350ms is in a H-mode configuration with \#24600 at 265ms being in L-mode. Both discharges are well diagnosed, having approximately 76 pickup coil, 24 flux loop, and 31 MSE observations recorded and available for equilibrium inference. Values for BEAST run parameters used in the following results are presented in \tabref{tab::beastRunParams}.

\begin{table}[!hbt]
\centering
\begin{tabular}{|l|c|}
\hline
\textbf{Run Parameter} & \textbf{Value}\\
\hline
\verb+sizeSamplePool+  & 150\\
\verb+numEvidenceSamples+ & 36\\
\verb+numABIFailures+ & 1000\\
\verb+numMCMCJumps+ & 20\\
\hline
\end{tabular}
\caption{Typical run-parameters for BEAST used in equilibrium inference for MAST discharges \#22254 at 350ms and \#24600 at 265ms. Definitions for each these run parameters are given in \tabref{tab::beastMinRunParams} in \appref{sec::Computation}.}
\label{tab::beastRunParams}
\end{table}

In addition to the toroidal/poloidal currents, currents for conducting surfaces and poloidal field coils were also inferred using the same Biot-Savart forward models detailed in \secref{sec::DiagnosticModels}. Moreover, additive biases to both pickup coil and flux loop signals were inferred to offset any errors in magnetic calibrations. These additional model parameters are treated as nuisance parameters in the final inference of plasma currents and kinetic pressure. The data in \tabref{tab::runTimes} in \appref{sec::Computation} corresponds to inferences run with these additional nuisance parameters. An avenue of research being explored is to characterise the impact of pre-computing these conducting surface currents and biases using the analytic current tomography introduced by Svenssion and Werner\cite{svensson2008,vonnessi2012} and subsequently locking these values in a non-analytic BEAST inference. The advantage of this would be speeding up BEAST execution times beyond the values reported in \tabref{tab::runTimes} in \appref{sec::Computation}.

The reader will note that BEAST approximates moments of the posterior (and marginalisations thereof) via Monte Carlo sampling statistics. To keep the text concise, the traditional terms for posterior moments (e.g. `expectation') will be used in this section but should be understood as Monte Carlo estimators thereof.

\subsection{Toroidal Current Density Inference}
\begin{figure}[!htb]
\includegraphics[width=\textwidth]{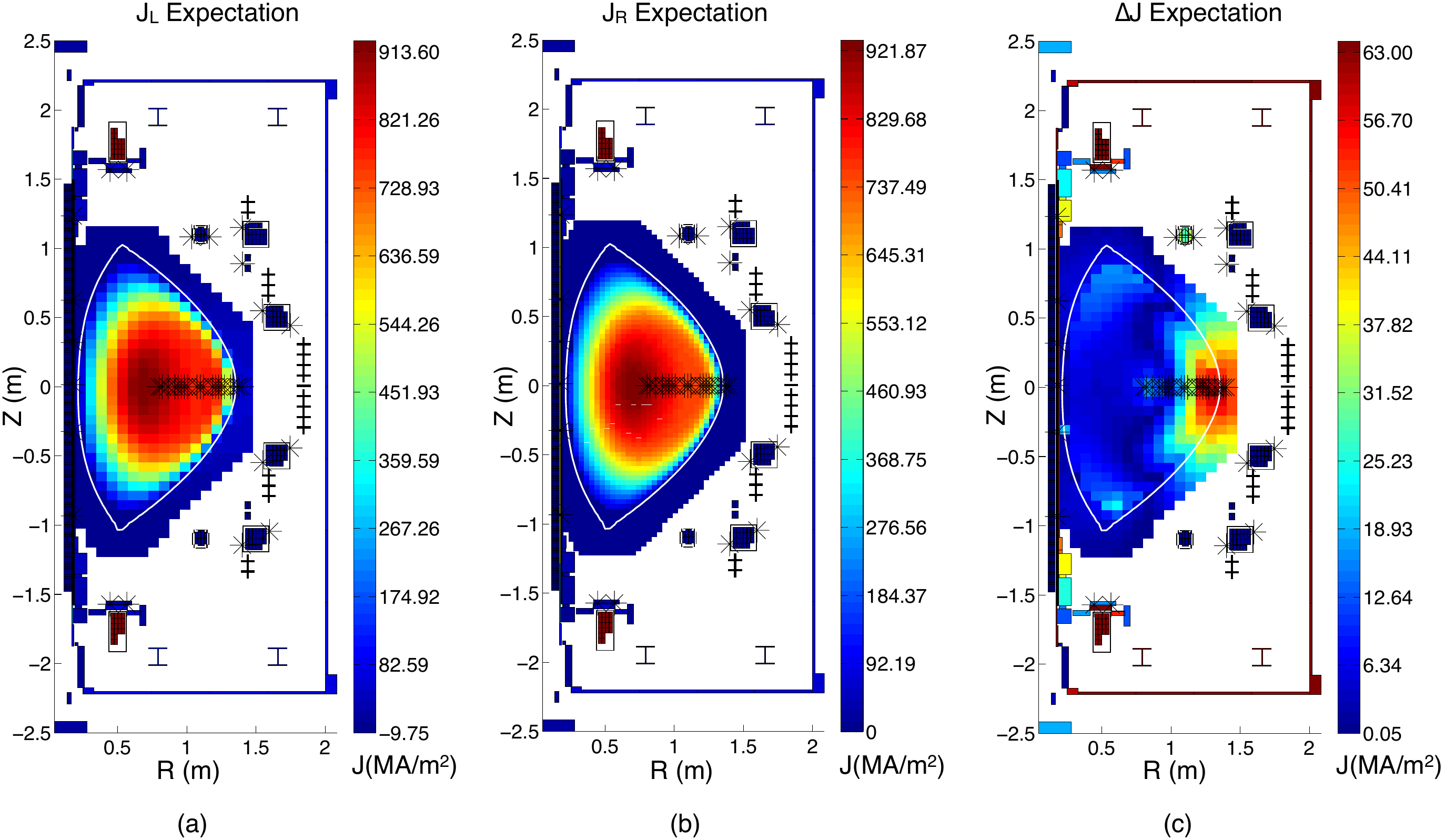}  
\caption{\label{fig::22254BeamData}Expectation values of $\JL(R,Z)$, $\JR(R,Z)$ and $\Delta J(R,Z)$ inferred for MAST discharge \#22254 at 350ms, as calculated from 1800 samples of the posterior, using pickup coils, flux loops, MSE and Rogowski coil data. The inferred LCFS is indicated in white on each figure. Flux loop locations are indicated by stars outside the plasma region; position and orientation of pickup coils are indicated via heavy bars on the out-board edge of the first wall and as a vertically oriented column line along the solenoid; and MSE observation positions are indicated by the stars across the mid-plane inside the plasma region. (a) shows $\JL(R,Z)$ current density data, with the current densities in (b) reflecting that of $\JR(R,Z)$. Note that the number and size of beams representing $\JL(R,Z)$ and $\JR(R,Z)$ are allowed to differ in BEAST inferences. (c) shows the magnitude of the current density difference as averaged across each 2D rectangular step corresponding to $\JL(R,Z)$.}
\end{figure}

\Figref{fig::22254BeamData} shows the expectations of $\JL(R,Z)$, $\JR(R,Z)$ and $\Delta J(R,Z)$ for \#22254 at 350ms, with an inferred $\sigma_*^2 = 9.461\times10^{-3}\pm1.2\times10^{-5}(kA)^2$, where the uncertainty reflects the 95\% confidence interval. These expectations have been calculated using sampling statistics generated from 1800 simulated samples of the posterior. The pixelation in \figref{fig::22254BeamData} (a) is simply a reflection of the plasma beam model described in \secref{sec::PlasmaModel} and displayed in \figref{fig::PlasmaBeam}. The pixelation in \figref{fig::22254BeamData} (b) reflects the fact that $\JR(R,Z)$ is approximated as a collection of densely packed, axisymmetric current beams, which enables the same set of algorithms to be used in forward model calculations associated with both $\JL(R,Z)$ and $\JR(R,Z)$. However, as $\JR(R,Z)$ itself is a derived structure from $\IL$, $\fc$ and $\pc$, increasing the number of beams used to represent this function (beyond just the number of $\IL$ model parameters) does not increase the degrees of freedom in the inference. Thus, a more dense set of beams was selected to represent $\JR(R,Z)$ to increase the accuracy in associated forward model calculations. Finally, one will note that the definition of $\Delta J(R,Z)$ in \eqref{eq::currentDiff1} is such that it will intrinsically reflect the beam configuration associated with $\IL$.

\Figref{fig::22254BeamData} (c) shows the regions of the plasma where the highest discrepancies from a GS solution are occurring. One will note that the $\Delta J(R,Z)$ values in \Figref{fig::22254BeamData} are relatively small, when compared to both $\JL(R,Z)$ and $\JR(R,Z)$. While $\Delta J(R,Z)$ can give some indication as to physical effects neglected in the GS equation, it can also be a reflection of where the plasma is simply more constrained by diagnostic observations. Thus, $\Delta J(R,Z)$ should be taken as a queue of how one may be able to resolve additional physics but can not be used to infer such physics directly (at least without employing additional constraints). In the case of \#22254 at 350ms, it is clear that the MSE observations are the source of the GS discrepancy seen on the outboard edge of the plasma.

\begin{figure}[!htb]
\includegraphics[width=\textwidth]{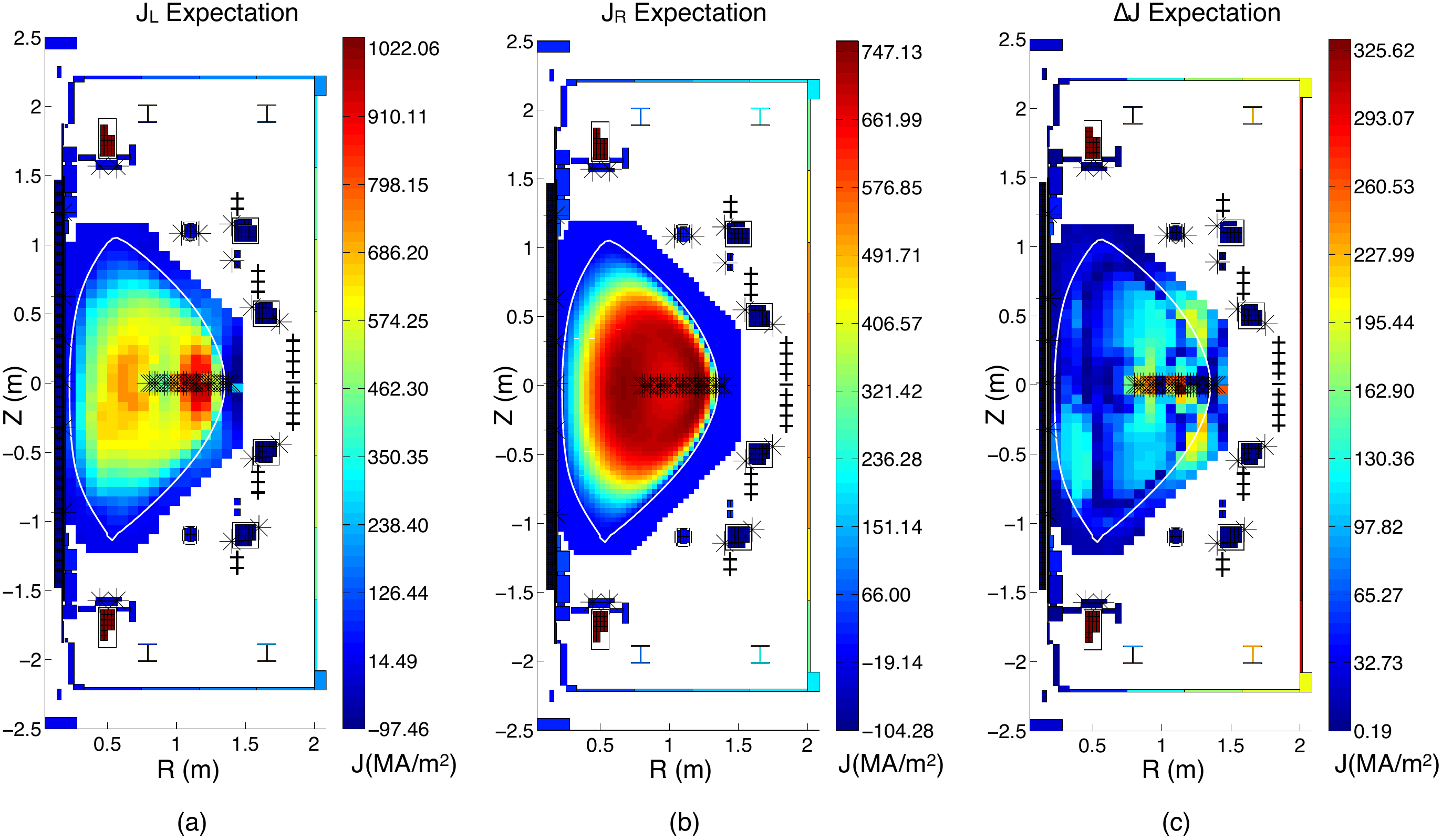}  
\caption{\label{fig::24600BeamData}Expectation values of $\JL(R,Z)$, $\JR(R,Z)$ and $\Delta J(R,Z)$ inferred for MAST discharge \#24600 at 265ms, calculated from 1800 posterior samples. These figures are inferred using the same diagnostics and have the same visualisation as in \figref{fig::22254BeamData}.}
\end{figure}

Toroidal current density data for discharge \#24600 at 265ms is presented in \figref{fig::24600BeamData}, which is analogous to the data for \#22254 at 350ms in \figref{fig::22254BeamData}. For the strength of the GS constraint, $\sigma_*^2 = 0.203381\pm1.59\times10^{-4}(kA)^2$, was inferred, which is substantially larger than in the case of \#22254 at 350ms. Corresponding to this, one can see that $\Delta J(R,Z)$ is also larger for discharge \#24600 than for \#22254 and show significant, localised deviations from a force-balance solution.

\begin{figure}[!htb]
\includegraphics[width=.667\textwidth]{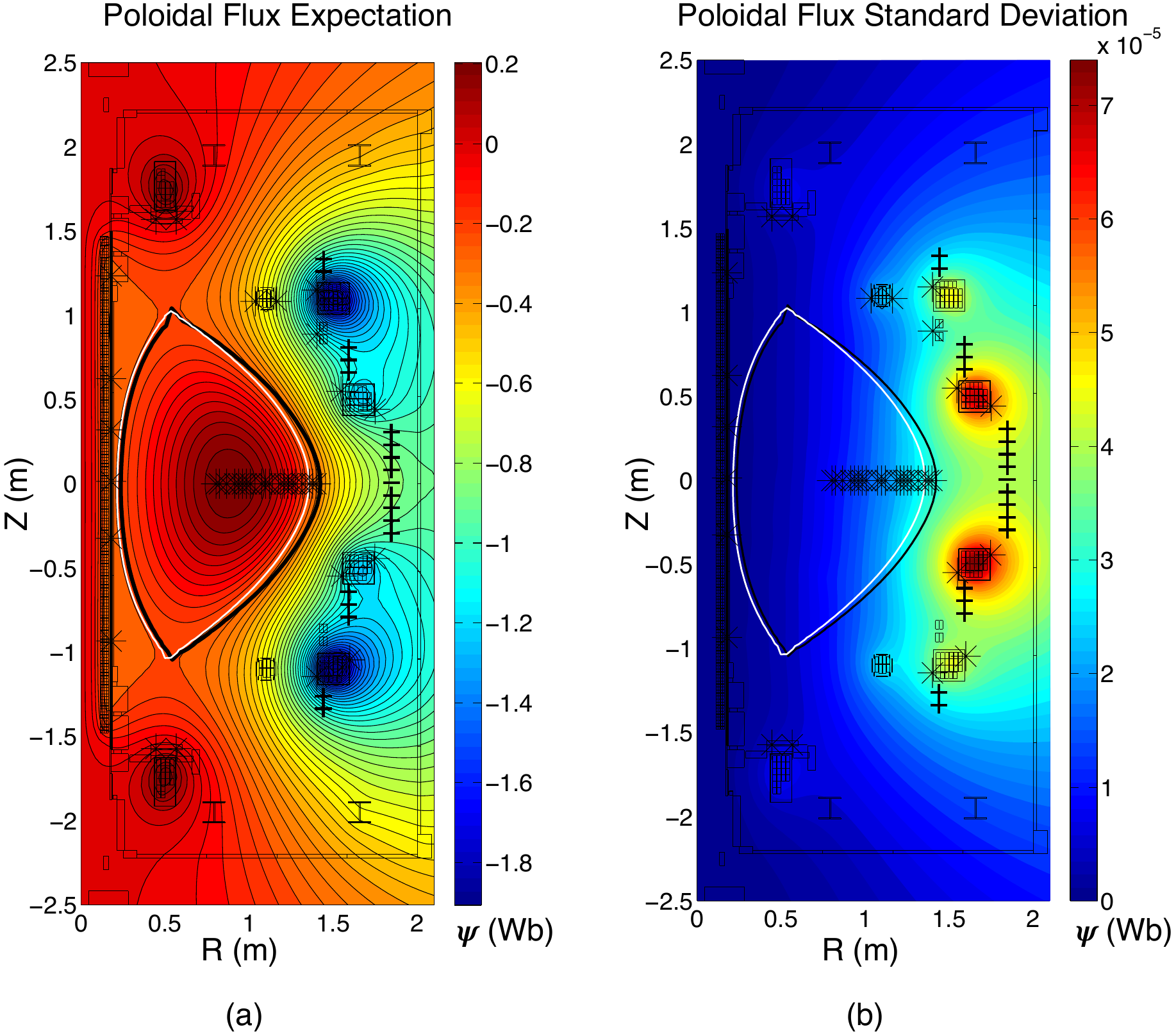}  
\caption{\label{fig::22254PoloidalFlux}Poloidal flux function expectation and standard deviation as calculated from 1800 samples of the posterior from shot \#22254 at 350ms. Positions of magnetics and MSE observation points are indicated as they were in \figref{fig::22254BeamData}. In (a) and (b) the EFIT LCFS is plotted in black with the inferred LCFS overlaid in white.}
\end{figure}

\subsection{Inference of the Poloidal Flux}
By direct application of Ampere's law, the poloidal flux function and statistical moments thereof can be calculated from samples of $\IL$. \Figref{fig::22254PoloidalFlux} shows the poloidal flux function expectation and standard deviation as calculated from 1800 samples of the posterior from \#22254 at 350ms. As the magnetic field geometry is very similar for discharge \#24600 at 280ms, the poloidal flux cross-section is only presented for \#22254.

\Figref{fig::22254PoloidalFlux} also presents comparisons between the LCFS as calculated by both BEAST and EFIT. The LCFS coming from both codes are in excellent agreement, especially around the X-point of the plasma. The discrepancy of the LCFS between BEAST and EFIT on the outboard edge is due to the presence of MSE observations making BEAST infer an LCFS which is slightly withdrawn into the plasma, when compared to EFIT, which was not utilising MSE constraints in this particular discharge.

\begin{figure}[h]
\includegraphics[width=\textwidth]{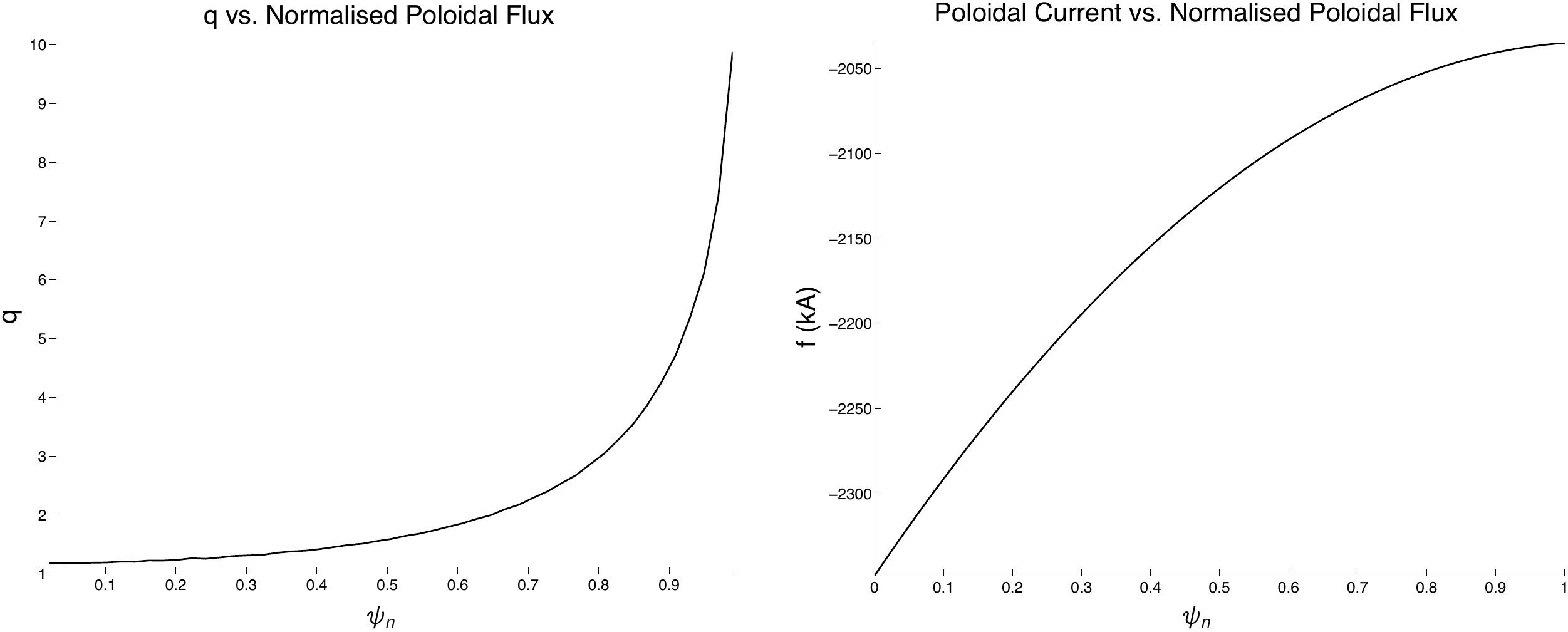}  
\caption{\label{fig::22254qfprofiles}Inferred $q$ and poloidal current profiles as a function of normalised poloidal flux for \#22254 at 350ms. The lines indicate expectations of the associated profile, with uncertainties suppressed, as they are too small to visually distinguish on the presented scales. MAP values for these quantities are also not plotted, as they too are not able to be visually distinguished from the expectation values.}
\end{figure}

\subsection{Profile Inference and Errors}
Statistical moments for $q$ and poloidal current are also routinely inferred in BEAST computations. Unlike the kinetic pressure, these quantities are directly constrained by the MSE observations and thus retain a standard physical interpretation. Indeed, the profiles presented in \figref{fig::22254qfprofiles} remained virtually unaltered when the forward models associated with $\JR(R,Z)$ were removed from the BEAST inference and the GS prior replaced by a GP smoothening prior (see von Nessi, et. al.\cite{vonnessi2012} about the use of Gaussian Processes in current tomography). As the shape of the poloidal current profile is strongly constrained by the MSE observations, via \eqref{eq::FAndPModels1a} and \eqref{eq::DiagnosticModels6}, with an intrinsic constraint to the toroidal field coil current (c.f. \eqref{eq::FAndPModels3}), it is expected that this profile would have little to no uncertainty in the inference. Moreover, as $q$ is closely related to the magnetic pitch angle, MSE measurements alone serve to strongly constrain this profile.

Beyond the above justifications, the models presented in this paper will generally lead to inference on equilibrium parameters with very small uncertainties. That is, the inference reflects a variational problem with a unique solution. As stated throughout the paper, BEAST infers solutions which are the most consistent with all diagnostic observations under a priori constraints imposed by the prior and forward models. In practice, degeneracies in model parameter configurations that lead to uncertainties in inferred quantities are greatly reduced when using \eqref{eq::Prior2} as part of the prior in conjunction with the weak observations in \secref{sec::ObservationSplitting}. Indeed, even when increasing the prior variance to $\sigma_*^2 = 100 (kA)^2$, uncertainties remain small in all inferred quantities. This behaviour was cross-verified by using both NS and Hybrid Markov Chain Monte Carlo (HMCMC) algorithms to extract samples from posterior distributions in BEAST inferences.

\begin{figure}[h]
\includegraphics[width=.75\textwidth]{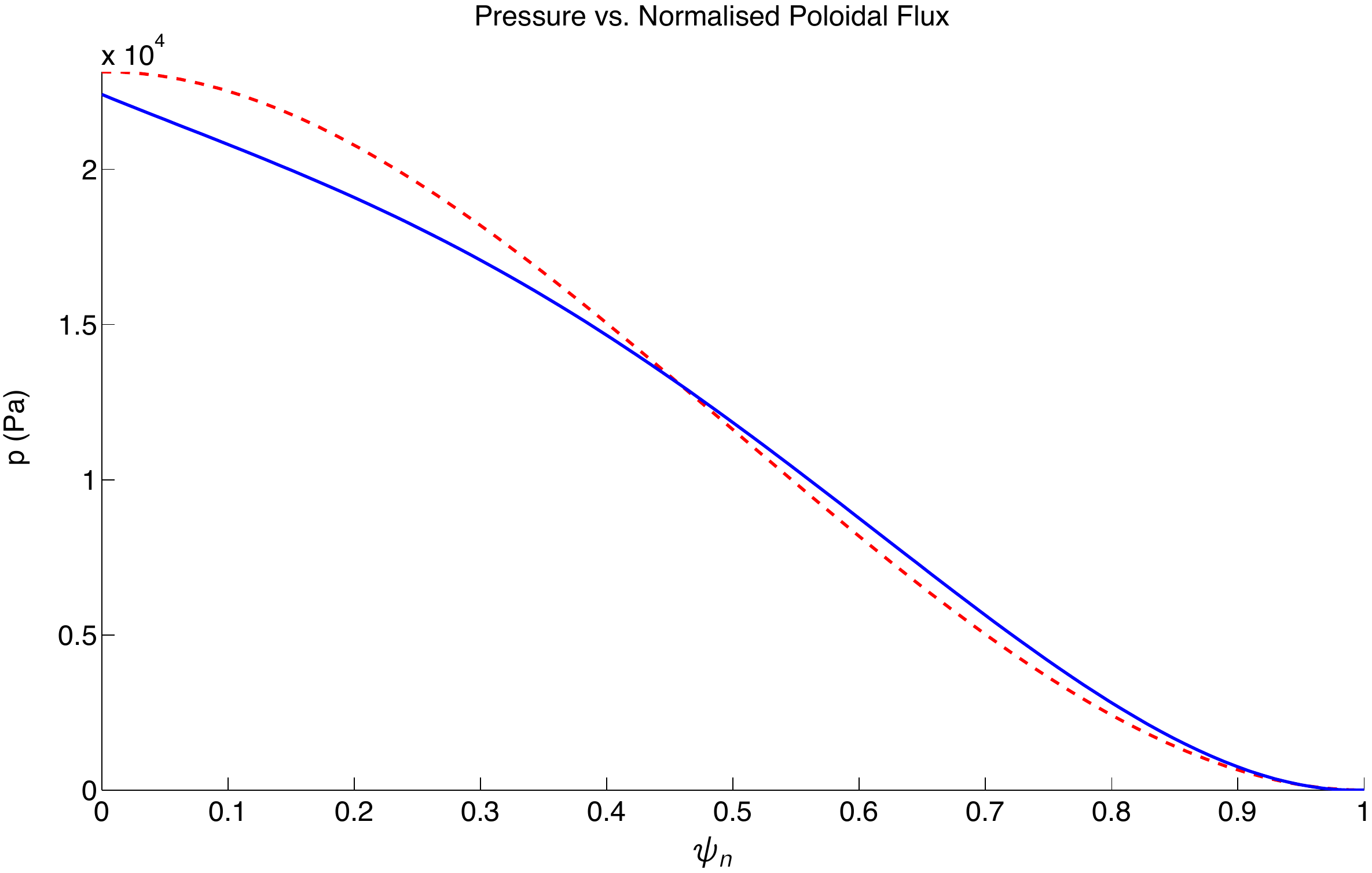}  
\caption{\label{fig::22254PressureComp}Inferred pressure profile as a function of normalised poloidal flux for \#22254 at 350ms. The blue line indicates the expectation of the pressure profile as calculated from BEAST.  The dashed, red line is the pressure profile calculated from EFIT. The comparison is only meant to indicate that the discrepancies from force-balance shown in \figref{fig::22254BeamData} (c) correspond to a physical pressure profile.}
\end{figure}

As discussed in \secref{sec::currentDiff}, pressure in BEAST inferences is effectively a nuisance parameter; but it is useful to verify that the kinetic pressure being inferred is still physically plausible. \Figref{fig::22254PressureComp} presents the inferred pressure profile for \#22254 as a function of normalised flux with a comparison to the EFIT-calculated pressure profile. This comparison is not meant to be any validation of either BEAST or EFIT beyond indicating that the discrepancies shown in \figref{fig::22254BeamData} (c) and \figref{fig::24600BeamData} (c) correspond to physically plausible pressure, as opposed to a vacuum or negative pressure solution to the GS equation. Indeed, without any direct diagnostic constraints on the pressure, no rigorous, physical interpretation of the BEAST-inferred pressure can be made.

\subsection{Evidence Calculations}

The NS sampling algorithm essentially preforms the posterior integration via a statistical quadrature and thus has intrinsic uncertainties (see \appref{sec::Computation} and/or Sivia and Skilling\cite{sivia2006} for details on this topic). For \#22254 at 350ms $\ln(P(D)) = 764.640\pm 1.039$; and for \#24600 at 265ms, $\ln(P(D)) = -39.439\pm1.033$, where the uncertainties are 95\% confidence intervals. These values correspond to the inferred $\sigma_*^2$ values for both discharges, in that \#24600 at 265ms has a significantly larger inferred value of $\sigma_*^2$ which reflects the fact that a larger degree of freedom was needed (relative to 22254 at 350ms) to accurately predict the diagnostic observations, which means that 24600 at 265ms has a smaller Bayes' factor\cite{mackay2003,sivia2006}. 

While results have been presented here for two MAST discharges, it should be noted that BEAST inferences have been run on dozens of discharges and run parameter configurations. Indeed, most inferences have produced data that strongly resembles what has already been presented in this section; and thus, have not been detailed further in this paper.

\section{Conclusions}\label{sec::Conclusions}
The feasibility of using Bayesian analysis in inferring plasma equilibria in a tokamak under a force-balance constraint has been demonstrated. In particular, the new technique of weak observation was introduced, which subsequently enabled GS solutions to be efficiently characterised in the prior and inferred differences from such solutions to be quantified. Moreover, an implementation of the NS algorithm has been presented which can be utilised as the foundation for high-dimensional, non-analytic inference problems. These two points have culminated in a code, named BEAST, which can not only infer equilibrium parameters but can also be an indicator of new physics by quantifying discrepancies from GS solutions directly and via the evidence associated with the inference itself. This code also has the capability of being readily modified to include new physics that amends GS force-balance and can subsequently be utilised to validate different force-balance models against one another via experimental data. Given these points, it is clear that this code exists as an ideal complement to the fast GS solvers already running routine analysis on many of today's tokamaks.

Current research endeavours surround exploiting BEAST as physics exploration tool in equilibrium studies on the MAST experiment. In particular, the construction of a direct and computationally tractable kinetic pressure constraint is currently being pursued. However, work is also being done to re-implement BEAST a parallelised code to run on a wide range of computational platforms, including HPCs, with an ultimate goal of making BEAST inferences an arbitrarily fast computation (depending on the number of available CPUs) suitable for routine post-analysis.

Finally, given the complexity and number of unknowns associated with equilibrium inference, there is a question surrounding the use and abilities of machine learning techniques in this field. However, the results from \secref{sec::Results} indicate that diagnostic observations can be accurately predicted by model configurations that are very close, in terms of $\sigma_*^2$, to a GS solution; the implication being that statistically trained empirical models will yield results very close to those coming from a standard code which fits GS solutions to diagnostic observations. As MAST is a well diagnosed experiment by today's standards, it is suspected that this situation is true for most tokamaks currently in operation. On the other hand, with the inclusion of newly-developed, highly-informative imaging diagnostics (e.g. the 2D MSE system developed by Howard\cite{howard2008}), significant deviations from GS solutions may well be inferred, even in standard operational scenarios. In this situation, it would then make sense to pursue development of empirically trained models to extend the abilities of today's GS solvers. Thus, BEAST represents a first step down the path of utilising machine learning techniques in tokamak equilibrium inference, as it utilise GS constraints while retaining the ability to quantify deviations from GS solutions.

\appendix

\section{Evidence Calculation and Posterior Analysis using Nested Sampling}\label{sec::Computation}
What follows is a review of Skilling's Nested Sampling (NS) algorithm and its implementation in BEAST. Specifically, a modified implementation of this algorithm is used to calculate the posterior evidence as well as simulate posterior sampling in BEAST inferences.

\subsection*{Transformation of the Evidence Integral}\label{sec::NestedSampling}
The NS algorithm was originally created to calculate evidence integrals for Baysian posteriors (c.f. the denominator in \eqref{eq1}), which were sufficiently complex and/or of high dimension such that they were poorly integrated via standard analytic or statistical quadratures. However, before being able to utilise NS statistics, the evidence integral first needs to be transformed. While this transformation is normally presented as part of the nested sampling algorithm\cite{skilling2006,sivia2006}, these presentations tend to be heuristic in nature. What follows is a concise, rigorous derivation of this transform which is then put into context of the overall algorithm used to calculate the evidence.

In essence, NS involves transforming the multi-dimensional evidence integral into a one-dimensional integral, which is amenable to statistical integration via the a statistical technique called ``nested sampling''. Taking $n$ to represent the dimension of the posterior domain, one can write
\begin{eqnarray}
\p(\vx,\vsigma) & = & \int_{\mathbb{R}^n} \p(\vx,\vsigma|\vm)\p(\vm)\,d\vm\nonumber\\
& = & \int_{\mathbb{R}^n}\int_0^{\p(\vx,\vsigma|\vm)}\p(\vm)\,dtd\vm\nonumber\\
& = & \int_0^\infty\int_{\left\{\vm\,|\,\p(\vx,\vsigma|\vm)>t\right\}}\p(\vm)\,d\vm dt.\label{eq::NestedSampling1}
\end{eqnarray}
Next, the following function can be defined:
\begin{eqnarray}
\xi(t) & := & \int_{\left\{\vm\,|\,\p(\vx,\vsigma|\vm)>t\right\}}\p(\vm)\,d\vm\nonumber\\
& = & \text{prior proportion with likelihood greater than }t.\label{eq::NestedSampling2}
\end{eqnarray}
It is clear that $\xi(t)$ is a decreasing function having range $[1,0]$ and domain $[0,\infty)$. Thus, $\xi^{-1}(t)$ is also decreasing and defined up to a set of measure zero on the domain $[0,1]$ (although not necessarily continuous). It is a classical result of real-analysis (c.f. Royden\cite{royden1988} Ch. 5 Thm. 3) that enables one to conclude that $\xi^{-1\prime}(t)$ exists almost everywhere in the domain and is Lebesgue integrable. Thus, we can rewrite \eqref{eq::NestedSampling1} using \eqref{eq::NestedSampling2}, make the substitution $v=\xi(t)$ and integrate by parts to find:
\begin{eqnarray}
\p(\vx,\vsigma) & = & \int_0^\infty\xi(t)\,dt\nonumber\\
& = & \int_1^0v\cdot\xi^{-1\prime}(v)\,dv\nonumber\\
& = & v\cdot\xi^{-1}(v)\Bigg|_1^0+\int_0^1\xi^{-1}(v)\,dv\nonumber\\
& = & \int_0^1\xi^{-1}(v)\,dv.\label{eq::NestedSampling3}
\end{eqnarray}
Unfortunately, the integral representation in \eqref{eq::NestedSampling3} is poorly approximated by standard quadratures as most of the integrand's mass is highly condensed around $v=0$. However, it will be shown below, that this representation is particularly amenable to integration via a statistical quadrature based on the nested sampling principle.

The closed form expression of $\xi^{-1}(t)$ is an integral over $n-1$ dimensions and offers little help in the evaluation of \eqref{eq::NestedSampling3}, as all the problems associated with the original integration re-emerge. Moreover, getting an intuitive idea of $\xi^{-1}(t)$ to understand how it can be statistically integrated is a subtle endeavour. However, one can start to get this understanding by considering a number, say $m$, of samples extracted from the prior ordered from highest to lowest associated likelihood value. Denoting $\mathcal{L}_i$ as the likelihood value of the $i$th ordered sample with $\mathcal{L}_1>\mathcal{L}_2>\cdots>\mathcal{L}_m$, one then has the following approximation:
\begin{equation}
\xi(\mathcal{L}_i)\approx\frac{i}{m}\Longrightarrow\mathcal{L}_i\approx\xi^{-1}\left(\frac{i}{m}\right).\label{eq::NestedSampling4}
\end{equation}
That is to say, if given $m$ prior samples ordered according to decreasing associated likelihood value, one expects that approximately $i/m$ of samples coming from the prior (i.e. proportion of the prior) will have associated likelihood values greater than $\mathcal{L}_i$. It is in this sense that likelihood-ordered prior samples can be associated with uniform samples of the abscissa for the integrand in \eqref{eq::NestedSampling3}. Indeed, using $t$ to denote a sample from a uniform distribution on $[0,1]$, if one were to take $m$ such samples and re-order them so that $t_m>t_{m-1}>\cdots>t_1$, then \eqref{eq::NestedSampling4} can be written as
\begin{equation}
\xi(\mathcal{L}_i)\approx t_i\Longrightarrow\mathcal{L}_i\approx\xi^{-1}\left(t_i\right).\label{eq::NestedSampling5}
\end{equation}
This suggests that one can integrate using quadratures associated with multiple resamplings of the abscissa (i.e. collections of $m$ uniform samples on $[0,1]$)  to find uncertainties in the evidence, which are a direct consequence of the approximation in \eqref{eq::NestedSampling5}. The specifics of the numerically integrating the expression in \eqref{eq::NestedSampling3} using \eqref{eq::NestedSampling5} will be discussed below.

\subsection*{Generating Posterior Moments via Sample Simulation}\label{sec::StaircaseSampling}
The values of $\xi^{-1}(t)$ used in the evidence integral calculation can also be used to simulate samples from the posterior. More specifically, with the evidence known, it is possible to associate a posterior probability with each prior sample used in the evidence quadrature. Sampling the values used to construct the $\xi^{-1}(t)$ graph according to their posterior probabilities then serves as a proxy (i.e. simulation) for direct sampling from the posterior. Thus, using these simulated samples, moments of any function of the posterior pdf can be calculated using simple sampling statistics.

To gain a rigorous understanding of the above statements, it is necessary to calculate the derivative of $\xi(t)$, which can be done via the co-area formula (c.f. Federer\cite{federer2006} Sec. 3.2) and Leibniz rule:
\begin{eqnarray}
\xi^\prime(t) & = & \frac{d}{dt}\int_{\left\{\vm\,|\,\p(\vx,\vsigma|\vm)>t\right\}}\p(\vm)\,d\vm\nonumber\\
& = & \frac{d}{dt}\int_t^\infty\int_{\left\{\vm\,|\,\p(\vx,\vsigma|\vm)=s\right\}}\frac{\p(\vm)}{|\nabla \p(\vx,\vsigma|\vm)|}\,d\mathcal{H}_{n-1}(\vm)ds\nonumber\\
& = &- \int_{\left\{\vm\,|\,\p(\vx,\vsigma|\vm)=t\right\}}\frac{\p(\vm)}{|\nabla \p(\vx,\vsigma|\vm)|}\,d\mathcal{H}_{n-1}(\vm),.\label{eq::StaircaseSampling1}
\end{eqnarray}
where $\mathcal{H}_{n-1}$ is the Hausdorff measure of co-dimension $1$ relative to the dimension of model parameters. Taking arbitrary $0<v_0<v_1<1$, one can now calculate
\begin{eqnarray}
\int_{v_0}^{v_1}\xi^{-1}(v)\,dv = \int_{\xi^{-1}(v_1)}^{\xi^{-1}(v_0)}t\int_{\left\{\vm\,|\,\p(\vx,\vsigma|\vm)=t\right\}}\frac{\p(\vm)}{|\nabla \p(\vx,\vsigma|\vm)|}\,d\mathcal{H}_{n-1}(\vm)dt,\label{eq::StaircaseSampling2}
\end{eqnarray}
via the substitution of $v = \xi(t)$ and using \eqref{eq::StaircaseSampling1}. On the other hand, the co-area formula and the definition of $\xi(t)$ also indicate
\begin{eqnarray}
& & \int_{\left\{\vm\,|\,\xi^{-1}(v_0)>\p(\vx,\vsigma|\vm)>\xi^{-1}(v_1)\right\}}\p(\vx,\vsigma|\vm)\p(\vm)\,d\vm = \nonumber\\
& &  \qquad \qquad\int_{\xi^{-1}(v_1)}^{\xi^{-1}(v_0)}t\int_{\left\{\vm\,|\,\p(\vx,\vsigma|\vm)=t\right\}}\frac{\p(\vm)}{|\nabla \p(\vx,\vsigma|\vm)|}\,d\mathcal{H}_{n-1}(\vm)dt.\label{eq::StaircaseSampling3}
\end{eqnarray}
Thus, equating \eqref{eq::StaircaseSampling2} and \eqref{eq::StaircaseSampling3} leads to
\begin{eqnarray}
\int_{v_0}^{v_1}\xi^{-1}(v)\,dv = \int_{\left\{\vm\,|\,\xi^{-1}(v_0)>\p(\vx,\vsigma|\vm)>\xi^{-1}(v_1)\right\}}\p(\vx,\vsigma|\vm)\p(\vm)\,d\vm.\label{eq::StaircaseSampling4}
\end{eqnarray}
As $\xi^{-1}(t)$ is a decreasing function, any partition of $[0,1]$ can be mapped to a finite covering of $\mathbb{R}^n$ via sets defined by $\left\{\vm\,|\,\xi^{-1}(v_i)>\p(\vx,\vsigma|\vm)>\xi^{-1}(v_{i+1})\right\}$, where $v_i$ and $v_{i+1}$ are successive points of an ordered partition on $[0,1]$. This cover will have pairwise non-intersecting members outside a set of measure zero, as $\xi^{-1}(t)$ is not necessarily monotone decreasing. Thus, \eqref{eq::StaircaseSampling4} allows for posterior probabilities to be associated with definite integrals of $\xi^{-1}(v)$. That is to say, if samples are drawn according to the posterior pdf and associated with values of $\xi^{-1}$ and $v$ (done by explicitly calculating the associated prior and likelihood probabilities of the sample and using the definition of $\xi(t)$), one will find that the number of samples within $[v_{i+1},v_i]$ (relative to other intervals) is precisely equal to the LHS integral in \eqref{eq::StaircaseSampling4}. Moreover, since the the full evidence integral can be calculated, one has that
\begin{eqnarray}
\p_{[v_0,v_1]}:=\int_{v_0}^{v_1}\frac{\xi^{-1}(v)}{\p(\vx,\vsigma)}\,dv,\label{eq::StaircaseSampling4b}
\end{eqnarray}
as being the proportion of posterior samples able to be associated with the interval $[v_0,v_1]$ for any $0\le v_0<v_1\le 1$. Relating this back to the evidence and \eqref{eq::NestedSampling5}, one can now write:
\begin{eqnarray}
\p(\vx,\vsigma) & = & \int_0^1\xi^{-1}(v)\,dv\nonumber\\
& = & \sum_{i=2}^m\int_{t_i}^{t_{i-1}}\xi^{-1}(v)\,dv\nonumber\\
& \approx & \sum_{i=2}^m\left[\mathcal{L}_i(t_{i-1}-t_{i})\right].\label{eq::StaircaseSampling4c}
\end{eqnarray}
Using \eqref{eq::StaircaseSampling4b} as the representation of the posterior probability for the integrals in the sum in line 2 of \eqref{eq::StaircaseSampling4c} enables one to associate the $i$th point of the evidence quadrature with a posterior probability:
\begin{equation}
\p_i:=\frac{\mathcal{L}_i(t_{i-1}-t_{i})}{\p(\vx,\vsigma)}\label{eq::StaircaseSampling5}
\end{equation}
That is to say, the prior sample used to generate the $i$th point of the $\xi^{-1}$ graph is taken to simulate a posterior sample with the posterior probability $\p_i$. 

\Eqref{eq::StaircaseSampling5} allows one to statistically calculate integrals of any function of the posterior pdf. In particular, the \emph{relative entropy}, denoted $\mathcal{E}$, between the posterior and prior can now be computed:
\begin{eqnarray}
\mathcal{E} & := & \int_{R^n} \p(\vm|\vx,\vsigma)\ln\left(\frac{\p(\vm|\vx,\vsigma)}{\p(\vm)}\right)\,d\vm\nonumber\\
& = & \int_{R^n} \p(\vm|\vx,\vsigma)\ln\left(\frac{\p(\vx|\vm,\vsigma)}{\p(\vx,\vsigma)}\right)\,d\vm\\
& \approx & \sum_i\p_i\ln\left(\frac{\mathcal{L}_i}{\p(\vx,\vsigma)}\right).\label{eq::StaircaseSampling6}
\end{eqnarray}
The relative entropy is a reflection of how much information is required to move from the prior to the posterior and will play a part in the termination criteria used by BEAST in its NS implementaiton, described below.

While this ability to calculate integrals/moments of the posterior is useful, it is often more convenient to exploit \eqref{eq::StaircaseSampling5} to simulate samples from the posterior directly. Indeed, \eqref{eq::StaircaseSampling5} enables posterior probabilities to be associated with the set of likelihood-ordered \emph{prior} samples used in \eqref{eq::NestedSampling5}. Thus, one is left with a set of model parameter configurations and associated posterior probabilities. This set can be viewed as a compressed version of the posterior, where extractions of model parameter configurations according to their respective posterior values are correctly interpreted as a simulation of direct sampling from the posterior. The caveat is that with every moment calculated from such a simulation, there will be associated uncertainties. However, this uncertainty has been consistently observed to be negligible in the context of the underlying errors associated with the results in \secref{sec::Results}. Thus, for the sake of clarity, only averages of these moments are presented in \secref{sec::Results}.

\subsection*{Numerical Integration via Nested Sampling}\label{sec::NumericalIntegration}

\Eqref{eq::NestedSampling5} already shows how a likelihood-ordered set of prior samples and sets of uniform samples on $[0,1]$ can be used to evaluate the integral in \eqref{eq::NestedSampling3}. In practice however, most of the mass associated with $\xi^{-1}(t)$ will usually be strongly condensed around $t=0$. In particular, an efficient abscissa spacing for evaluating the evidence, as expressed in \eqref{eq::NestedSampling3}, can vary over many orders of magnitude, when considering the whole domain of integration. Moreover, it's not possible to know the finest spacing needed to evaluation the integral to within a certain accuracy \emph{a priori}; this would require detailed knowledge of the posterior before any inference calculations were made. Even if one did know this number, it would normally imply the use of so many points in the quadrature that the calculation of the integral over most of the domain would be rendered inefficient to the point where the overall integration becomes computationally intractable due to requiring too much time and/or memory to calculate. Thus, naive analytic and statistical quadrature methods will still poorly approximate the integral in \eqref{eq::NestedSampling3}, due to $\xi^{-1}(t)$ needing a high and variable precision in the abscissa around $t=0$. 

Skilling\cite{skilling2006} developed the method of nested sampling to deal with the specific issue discussed above, which can be summarised in the following pseudo-code: 

\begin{enumerate}
\item Start with a set of $m$ samples from the prior, with a corresponding set of $m$ uniform, abscissa samples on $[0,1]$.
\item Order the prior samples according to their likelihood values, while ordering the abscissa samples according to their value.
\item Extract the prior sample with the lowest likelihood along with the abscissa sample with the highest value. Store this as a tuple $(\mathcal{L}_i,t_i)$, containing the likelihood and abscissa values respectively. Note that $i$ is meant to indicate the index of the extracted tuple.
\item Replenish the initial prior sample pool by continuing to extract a new prior sample until such is found with a likelihood greater than $\mathcal{L}_i$. Replenish the pool of abscissa samples by drawing a sample from a uniform distribution on $[0,t_i]$.
\item Using all extracted tuples construct a current approximation of the evidence and relative entropy using a trapezoidal quadrature, \eqref{eq::NestedSampling3}, \eqref{eq::NestedSampling5} and \eqref{eq::StaircaseSampling6}.
\item If the number of iterations exceeds $2m\mathcal{E}$, terminate the algorithm, else return to step 2.
\end{enumerate}

Note that without a priori knowledge of the posterior, the choice of termination criteria is subjective.\cite{sivia2006} The above pseudo-code reflects the termination criteria used in BEAST and is the one suggested in Sivia \& Skilling\cite{sivia2006}. For details regarding the expected accuracy and reasoning behind the choice to limit iterations to $2m\mathcal{E}$, the interested reader is encouraged to read section 9.2.2 of Sivia \& Skilling\cite{sivia2006}.

It is clear that the number $m$ elements will always be preserved in the prior and abscissa sample pools, while the likelihood/abscissa constraint will monotonically increase/decrease. This leads to a statistical quadrature that naturally accumulates around $t=0$, which solves the problem outlined in the previous section. While it is a simple matter to extract a uniform sample from $[0,t_i]$ for any $t_i>0$, extracting prior samples under a growing likelihood constraint will become an increasingly difficult computation as the iteration proceeds. In the context of BEAST inference, sampling the prior \emph{ab initio} (i.e. naively sampling the prior and then testing its likelihood), will quickly have the algorithm failing an unacceptable amount of times before generating a prior sample that can meet the current likelihood constraint. Thus, a method was developed that would continue to efficiently generate prior samples throughout all iterations of the algorithm. This technique is outlined in the next section.

Finally, as discussed earlier, multiple abscissa sample sequences are generated to calculated the uncertainties of the evidence using sampling statistics. As nested sampling of both the prior and abscissa samples can proceed independently, BEAST works by generating one sequence of likelihood-ordered prior samples and many sequences of abscissa samples to calculate samples of the evidence integral. Once the first abscissa sequence is generated, the above pseudo-code is rerun using stored likelihood-ordered prior samples with a new abscissa being generated as the iteration proceeds. The prior sample sequence is extended in the case where more prior samples are needed to meet the termination criteria. In practice rerunning the above pseudo-code with stored prior values takes a very small fraction of the time, as compared to when the prior sample sequence has to be actively generated. Once the above iteration has run for all abscissa sequences, sampling statistics are applied to the calculated evidences for each abscissa.

\subsection*{Priori Sampling and Optimal Seeding}\label{sec::OptimalSeeding}
Unfortunately, in the high dimensional inferences associated with BEAST, NS is still not sufficient to accurately calculate the evidence of the posterior in a time which would be useful to scientists (more than weeks for a typical equilibrium inference as outlined in this paper). Indeed, typical posteriors for BEAST inferences have highly-localised probability densities, with many local maxima dispersed throughout the model parameter domain. As NS is fundamentally based on a statistical quadrature, it will typically require exceedingly long times to 'find' regions where the majority of posterior mass resides. Moreover, generating prior samples becomes increasingly difficult as the likelihood constraint increases, in that \emph{ab initio} sampling will quickly reach a point where the vast majority of generated samples will fail to meet the likelihood constraint and thus be discarded.

To overcome the above issues, a special implementation of nested sampling was developed for BEAST. Generally, this method uses externally calculated local maxima of the posterior as potential starting points of Markov Chain Monte Carlo (MCMC) iterations to efficiently generate prior samples, regardless of the current likelihood constraint. Given that the MCMC iterations are partially seeded by local maxima, we call this method of prior sampling \emph{optimal seeding}. The prior sampling procedure is as follows:

\begin{enumerate}
\item In addition to the pool of initial prior samples, a collection of local maxima of the posterior are calculated (using standard optimisation algorithms) and stored, this extended set of prior samples and local maxima is denoted $\mathcal{S}$;
\item \emph{ab initio} sampling of the prior proceeds in the nested sampling iteration until a fixed number (in the results below this is taken to be 1000) of successive samples fail to meet the current likelihood constraint for a single attempt to replenish the prior sample pool. Subsequent prior samplings will immediately use MCMC iteration instead of re-attempting \emph{ab initio} sampling of the prior.
\item Once the above threshold is reached, an adaptive MCMC iteration is used to generate new prior samples. This MCMC iteration takes a random member from $\mathcal{S}$, as its starting point. The jump distribution for the MCMC chain is a Gaussian with dimensional standard deviations corresponding to the dimensional lengths of a minimal volume hypercube bounding $\mathcal{S}$. This jump distribution has its associated variance  scaled further--using acceptance rate data from chains in previous prior sample generations--to help ensure an optimal acceptance rate of 23.4\%. This rate ensures optimal sampling efficiency for the chain.\cite{roberts1997} The chain itself is designed to sample from a distribution proportional to the following:
\begin{equation}
\left\{\begin{array}{rl} \p(\vm),&\ \p(x_i,\sigma_i|\vm)>\mathcal{L}_i\\0,&\ \text{otherwise}\end{array}\right.,
\end{equation}
where $\mathcal{L}_i$ represents the likelihood constraint for the present iteration.
\item If the current likelihood constraint exceeds the associated likelihood value of any member of $\mathcal{S}$, this member is removed, i.e. members that were included in $\mathcal{S}$ as local maxima of the posterior can be removed under this condition.
\end{enumerate}
As the local maxima in $\mathcal{S}$ will generally have large likelihood values, these members will normally be potential starting points for prior MCMC chains after many iterations of nested sampling. In practice, optimal seeds are generated through a combination of conjugate gradient, Hooke/Jeeves and particle swarm optimisers (see Hassan, et. al. \cite{hassan2005} for details on the particle swarm heuristic). These seeds will typically be discarded after $m\mathcal{E}$ iterations, i.e. about half-way through the evidence calculation. As the bounding hypercube of the samples decreases (or will after some point) as the procedure continues, the jump distribution associated with the MCMC chains will become more resolved. This enables efficient sampling of the prior regardless of the likelihood constraint. Moreover, since $\mathcal{S}$ contains any number of local maxima, the algorithm will not waste time searching for regions of high probability density. 

This scheme provides a non-local approximation that captures the ambient structure of the posterior, as well as the finer structure around the local maxima included in $\mathcal{S}$. Finally, as $\mathcal{S}$ represent only the starting points of the MCMC iterations, they do not serve to bias the sampling of the prior, if a sufficient number of jumps are taken in the MCMC iteration itself. In BEAST, MCMC chains are fixed to make twenty jump attempts, if fewer than three jumps actually occur (i.e. the proposal state is accepted), the current chain is discarded and a new MCMC is started from a new random starting point taken from $\mathcal{S}$.

Finally, it should be noted that the issue of MCMC 'burn-in' is handled in the current implementation by running many short-length MCMC chains from a large number of initial starting points. This selection is a canonical one for NS, as the algorithm intrinsically stores a potentially-large number of samples at any given time. Details of this approach to handling MCMC burn-in, as well as others, can be found in Chapter 29 of MacKay\cite{mackay2003}.

 \subsection*{Benchmarks}\label{sec::Benchmarks}
 \Tabref{tab::beastMinRunParams} contains a list of run parameters which can be set in a BEAST inference, along with the suggested minimal values of each associated parameter. Smaller values for each parameter lead to overall faster computation, at the cost of consistency or inferred results. The values in \tabref{tab::beastMinRunParams} have been empirically observed to produce inferred results which are consistent with inferred uncertainties across different BEAST runs on the same MAST discharge using the same run parameters. However, inferences presented in \secref{sec::Results} use higher run parameter values (c.f. \tabref{tab::beastRunParams}) to get better statistics on final inferences.
 
 \begin{table}[!hbt]
\centering
\begin{tabular}{|l|c|c|}
\hline
\textbf{Run Parameter} & Variable Name & \textbf{Minimal values}\\
\hline 
\# of Prior Samples & \verb+sizeSamplePool+ & 20\\
\# of Evidence Abscissa Samples & \verb+numEvidenceSamples+ & 12\\
\# of \emph{ab initio} failures before using MCMC & \verb+numABIFailures+ & 0\\
\# of attempted jumps for prior MCMC iterations & \verb+numMCMCJumps+ & 12\\
\hline
\end{tabular}
\caption{Minimal run parameter values suggested for BEAST inference on MAST discharges.}
\label{tab::beastMinRunParams}
\end{table}

The performance of BEAST most strongly depends on the number of prior samples one wishes to maintain in nested sampling, with the other run parameters having little to no discernible impact on performance or results when set above the corresponding values in \tabref{tab::beastMinRunParams}. Specifically, to gain a stable uncertainty for the evidence, at least 12 samples should be taken of the evidence abscissa. One need not engage in naive sampling of the posterior at all and can simply use MCMC iteration for all prior sampling. However, it was seen that allowing for up to 1000 failures in naive prior sampling afforded the MCMC iterations to start off with much more resolved jump distributions, which subsequently made the overall evidence calculation time more consistent. While going below 20 MCMC jump attempts can make the overall computation faster, this was not heavily explored as it was desirable to have good assurance that the information contained in the MCMC starting point had sufficient time to dissipate (i.e. over the course of approximately $5$ jumps) at the optimal acceptance rate of 23.4\%.

Skilling\cite{sivia2006,skilling2006} discusses the impact that the initial number of prior samples has on the evidence uncertainty, taking $2m\mathcal{E}$ as the termination criteria. Thus, a detailed discussion on this topic here will not be presented here. \Tabref{tab::runTimes} gives an account of how the number of initial prior samples affected the time to calculate the evidence integral and what the computed uncertainties where for each of these calculations. Note that typically the evidence of the posterior is so large, that results from BEAST are actually output in terms of the natural logarithm of the evidence (log-evidence). One should note that each sample generated in column three of the table corresponds to one MCMC chain running and performing at least 20 evaluations of the posterior function, hence the long execution times. Speeding up sample generation via parallelisation of the posterior function evaluation is a current research pursuit.

While comparisons of the evidence between different BEAST runs with the same number of prior samples all produced results consistent with the reported uncertainties, using different numbers of starting prior samples will, generally showed an impact beyond the stated uncertainties. This impact was seen to be up to an order of magnitude above the stated uncertainties for the log-evidence. Regardless, this fluctuation is still small compared to the mean expectation of the log-evidence typically seen in a MAST discharge. More specifically, for MAST discharges, the log-evidence of the posterior will be calculated to normally be somewhere between 500 and 1000 with initial prior sample numbers causing fluctuations that are on the order of 10. As BEAST comparisons are completed using the same number of initial prior samples, relative comparison of the evidence remains meaningful in the context discussed in \secref{sec::Interpretation}. An overall fluctuation of the evidence across different starting numbers of prior samples is not surprising, as the termination criteria directly depends on this number. An attempt was made to see if there was a lower bound on the number of prior samples that would stabilise the evidence relative to bigger initial sampling pools; but this could not be determined, as some discharges had the evidence fluctuating up through an initial number of samplings that required more memory than was currently available. The search for this lower bound and the exploration of alternative termination criteria remains a focus of current research.

\begin{table}[!hbt]
\centering
\begin{tabular}{|c|c|c|c|c|}
\hline
\parbox{.19\textwidth}{\textbf{Number of Prior Samples}} & \parbox{.19\textwidth}{\textbf{Time to Compute Evidence (s)}} & \parbox{.19\textwidth}{\textbf{Nubmer of Quadrature Points}} & \parbox{.19\textwidth}{\textbf{Log-Evidence Expectation}} & \parbox{.19\textwidth}{\textbf{Log-Evidence 2$\sigma$}}\\
\hline
400 & 19811.737 & 37401 & 765.32 & 0.520 \\
200 & 9383.223 & 18512 & 770.28 & 1.022\\
150 & 6505.549 & 14322 &  764.64 & 1.039\\
100 & 4872.322 & 9211 & 777.21 & 1.154\\
50 & 2001.781 & 5029 & 750.11 &1.202\\
\hline
\end{tabular}
\caption{BEAST run times and evidence uncertainties for different numbers of initial prior samples using 25 abscissa samples. These statistics corresponds to BEAST analyse of MAST discharge \#22254 at 350ms.}
\label{tab::runTimes}
\end{table}

Finally, the data in \tabref{tab::runTimes} corresponds to BEAST implemented as a module in the MINERVA: a single-threaded, Bayesian java framework developed by Svensson\cite{svensson2007}. The hardware used to run BEAST for these tests was an iMac desktop PC running OS X 10.6.8 with a quad-core i7 processor clocking at 2.93GHz per core and having 8GB of memory. It should be noted that parallelising the NS algorithm is conceptually a straight-forward matter and would greatly reduce these computation times. This is a current research endeavour.

\begin{acknowledgments}
This work was jointly funded by the Australian Government through International Science Linkages Grant CG130047, the Australian Research Council Grant FT0991899, the Australian National University, the United Kingdom Engineering and Physical Sciences Research Council under grant EP/G003955, and by the European Communities under the contract of Association between EURATOM and CCFE. The views and opinions expressed herein do not necessarily reflect those of the European Commission.
\end{acknowledgments}

\bibliographystyle{unsrt}
\bibliography{beastRefs}

\end{document}